\documentstyle[11pt,amstex,righttag]{article}
\addtocounter{secnumdepth}{1}
\newcounter{exerc}
\numberwithin{equation}{section}

\begin{document}

\pagestyle{empty}
\title{Reply to\\ ``Comment on
`Critical behavior of a two-species\\
reaction-diffusion problem' ''\\}
 
\author{J.E. de Freitas$^1$, L.S. Lucena$^1$, L.R. da Silva$^1$,
and H.J. Hilhorst$^2$\\[2mm]
\begin{small}
$^1$Departamento de F\'\i sica Te\'orica e Experimental
\end{small}\\
{\small Universidade Federal do Rio Grande do Norte}\\
{\small Campus Universit\'ario, 59072-970 Natal, Brazil}\\[2mm]
{\small $^2$Laboratoire de Physique Th\'eorique$^*$}\\
{\small B\^atiment 210, Universit\'e de Paris-Sud}\\
{\small 91405 Orsay Cedex, France}\\}

\maketitle
\vspace{-1cm}
\begin{small}
\begin{abstract}
\noindent Recent Monte Carlo results [{\it Phys. Rev. E} {\bf 61} (2000)
6330] for the one-dimensional
reaction--diffusion process $A+B\to 2B$ and $B\to A$ lead to the
correlation length exponent estimate $\nu=2.21\pm 0.05$.
In a comment on our work 
[{\it cond-mat/0007366}] the exact value is claimed to be $\nu=2$. 
We reply that the arguments advanced fail to substantiate this claim.
\end{abstract}
\end{small}
\vspace{75mm}

\noindent L.P.T. - ORSAY 00/xx\\
{\small$^*$Unit\'e Mixte de Recherche du Centre National de la
Recherche Scientifique \- - UMR 8627}
\newpage
%%%%%%%%%%%%%%%%%%%%%%%%%%%%%%%%%%%%%%%%%

We recently considered \cite{FLSH} the two-species
reaction--diffusion process $A+B\to 2B$ and $B\to A$,  
which conserves the total particle density $\rho$.
At a critical density
$\rho=\rho_c$ there appears a phase transition 
from a stationary state with only the $A$ species present to
one with a nonzero $B$ density.
When $\rho\to\rho_c$ the
correlation length is believed to diverge as a power law 
with an exponent $\nu$.
On the basis of Monte Carlo simulations we estimated, among other things, 
that $\nu=2.21\pm 0.05$.
In a comment on our work Janssen \cite{Janssen00} claims that
the exact value is $\nu=2$.

The author's argument invokes 
material from several papers \cite{KSS,WOH,LHOW}, 
some coauthored by one of ourselves. 
It exploits the symmetries of a {\it truncated}
action and the fact that 
perturbatively in $\epsilon=4-d$
the truncated terms, which are quartic in the fields \cite{WOH},
are irrelevant under renormalization.
It then leads to $\nu=2/d$ in dimension $d$, valid to all orders in
$\epsilon$. 

The argument is of a well-known kind.
It deserves to be advanced, but it is certainly not decisive. 
The reason for that is simple and also well-known.
We simulated the {\it full} action \cite{FLSH}, which due to the quartic terms
breaks the relevant symmetries; these symmetry breaking terms 
become na\"\i vely relevant below dimension $d=2$. Unless it can be shown
how they fare under renormalization, the argument of Ref.\,\cite{Janssen00}  
is very different from a proof. The value $\nu=2$ is exact only
if one believes an unverified assumption.

Therefore our Monte Carlo results and the 
theoretical arguments in favor of $\nu=2$
will have to coexist awaiting
further investigation.
The present Monte Carlo data, in the absence of further
theoretical guidance,
at best provide some rough idea of how large the
corrections to scaling would have to be if $\nu$ were equal to 2.
On the analytical side a careful analysis of the symmetry breaking 
terms is required before any firm statements can be made.
\vspace{-3mm}

\section*{Acknowledgments}
\vspace{-1mm}

The authors acknowledge discussion with 
F. van Wijland.
This work is part of the French-Brazilian scientific
cooperation project CAPES-COFECUB 229/97. 
The authors also thank CNPq and Projeto Nordeste
de Pesquisa for support.
\vspace{-3mm}

%%%%%%%%%%%%%%%%%%%%%%%%%%%%%%%%%%%%%%%%%%%%%%%%%%%%%%%%%%%%%%%%%%%%%%%%%%%%%%%

\end{document}